\documentclass[12pt,english]{article}
\usepackage[a4paper, portrait,
	left=0.75in,
	right=0.75in,
	top=1in,
	bottom=0.5in,
	footskip=.25in]{geometry}
\usepackage[english]{babel}
\usepackage[utf8]{inputenc}
\usepackage[T1]{fontenc}
\usepackage{helvet,etoolbox,graphicx,titlesec}
\usepackage{caption,subcaption}
\usepackage[affil-it]{authblk} 
\usepackage{etoolbox}
\usepackage{lmodern}

\usepackage{mathrsfs}
\usepackage{amsfonts}
\usepackage{amsmath} 
\usepackage{amssymb} 

\usepackage{titlesec}

\usepackage{mathrsfs}

\usepackage{graphicx}
\usepackage{dcolumn}
\usepackage{bm}

\usepackage{amsthm}
\usepackage{xspace}
\usepackage[mathcal]{euscript}
\usepackage{tikz}
\usepackage{ifthen}

\usepackage{pgfplots}
\usepgfplotslibrary{patchplots,colormaps}
\pgfplotsset{width=7cm,compat=1.18,
colormap={mycolormap}{color=(black) color=(black!20!white)}}
\usepgfplotslibrary{fillbetween} 
\usepgflibrary{shadings}
\usetikzlibrary{backgrounds}

\usepackage{epsdice}
\usepackage{twemojis}

\usepackage{pifont}

\usepackage{tcolorbox}
\tcbuselibrary{theorems,skins,breakable}
\usepackage{float}
\usepackage{colortbl}

\usepackage{natbib}

\usepackage{orcidlink}

\usepackage{url}
\usepackage{hyperref}
\usepackage{color}
\definecolor{refcolor}{RGB}{160,35,0}
\definecolor{hrefcolor}{RGB}{0,35,190}
\hypersetup{
    colorlinks,
    citecolor=refcolor,
    filecolor=refcolor,
    linkcolor=hrefcolor,
    urlcolor=hrefcolor
}

\usepackage{booktabs}

\usepackage{bookmark}
\bookmarksetup{
  numbered, 
  open,
}

\theoremstyle{remark}

\numberwithin{proofStep}{theorem} 

\theoremstyle{definition}
\newtheorem{defCustom}{Definition of}
\renewcommand{\thedefCustom}{\arabic{definition}}
\makeatletter
\newcommand{\setdefCustomtag}[1]{
  \let\oldthedefCustom\thedefCustom
  \renewcommand{\thedefCustom}{#1}
  \g@addto@macro\enddefCustom{
    \global\let\thedefCustom\oldthedefCustom}
  }
\makeatother

\theoremstyle{definition}

\theoremstyle{remark}

\newcommand{\boxRule}{0.15mm}

\newcommand{\boxIndent}{15pt}

\newcommand{\remColor}{green}
\newcommand{\quoteColor}{blue}
\newcommand{\questColor}{red}
\newcommand{\figColor}{orange}
\newcommand{\tabColor}{orange}
\newcommand{\abstractColor}{blue!80!cyan}

\newenvironment{frameEnv}[1]
	{\begin{tcolorbox}[breakable,enhanced,toprule at break=0pt,bottomrule at break=0pt,before skip balanced=0.3cm,boxrule=\boxRule,left=0.75mm,right=0.75mm,frame hidden,borderline north = {\boxRule}{0pt}{#1!50!black}, borderline south = {\boxRule}{0pt}{#1!50!black},arc=0mm,colframe=#1!50!black,colback=#1!10,before upper={\parindent\boxIndent}]}
	{\end{tcolorbox}}

\newenvironment{frem}
	{\begin{frameEnv}{\remColor}}
	{\end{frameEnv}}
\newenvironment{fquest}
	{\begin{frameEnv}{\questColor}}
	{\end{frameEnv}}
\newenvironment{fquote}
	{\begin{frameEnv}{\quoteColor}}
	{\end{frameEnv}}
\newenvironment{ffig}
	{\begin{frameEnv}{\figColor}}
	{\end{frameEnv}}
\newenvironment{ftab}
	{\begin{frameEnv}{\tabColor}}
	{\end{frameEnv}}

	\newenvironment{frameEnvMargin}[1]
	{\begin{tcolorbox}[breakable,enhanced,toprule at break=0pt,bottomrule at break=0pt,before skip balanced=0.3cm,boxrule=\boxRule,left=0.75mm,right=0.75mm,top=5mm,bottom=5mm,frame hidden, borderline north = {\boxRule}{0pt}{#1!50!black}, borderline south = {\boxRule}{0pt}{#1!50!black},arc=0mm,colframe=#1!50!black,colback=#1!10,before upper={\parindent\boxIndent}]}
	{\end{tcolorbox}}

\newenvironment{fabstract}
	{\begin{frameEnvMargin}{\abstractColor}\begin{abstract}}
	{\end{abstract}\end{frameEnvMargin}}

\newcommand{\orcid}[1]{\href{https://orcid.org/#1}{\textcolor[HTML]{A6CE39}{\aiOrcid}}}

\def\({\left(}
\def\){\right)}

\newcommand{\mc}[1]{\mathcal{#1}}

\newcommand{\R}{\mathbb{R}}

\newcommand{\etc}{\textit{etc}}

\newcommand{\uu}{\mathfrak{u}}

\newcommand{\schrod}{Schr\"odinger}

\newcommand{\ket}[1]{|#1\rangle}

\newcommand{\x}{\mathbf{x}}

\newcommand{\emoji}[1]{\raisebox{-2.5pt}{\scalebox{1.5}{\twemoji{#1}}}}

\newcommand{\cmark}{\ding{51}}%
\newcommand{\xmark}{\ding{55}}%

\def\sref #1{\S\ref{#1}}

\def\ra{1.5}
\def\rb{1}
\def\rw{0.075}
\def\rt{0.35}
\pgfmathdeclarefunction{paramFuncX}{2}
{%
    \pgfmathparse{1.25*(#2*\ra*cos(#1 r)+#2*\rb*#1^\rt*sin(#1 r))}%
}%
\pgfmathdeclarefunction{paramFuncY}{2}
{%
    \pgfmathparse{0.7*(#2*\rb*sin(#1 r)-#2*\ra*#1^\rt*cos(#1 r))-0.5}%
}%

\newcommand{\classicalManyToMany}[8]{
\begin{tikzpicture}[line cap=round,scale=1.6]
\draw[help lines,cyan] (-4.2,-0.75) grid [step=1.0] (2.7,1.9);
\draw[-latex] (-4.25,0) -- (3.0,0) node [below] {$q$};
\draw[-latex] (0,-0.75) -- (0,2.0) node [left] {$p$};
\foreach[count=\j]\i in {-4,-3,-2,-1,1,2}
  \draw (\i,0.05) -- (\i,-0.05) node [below,gray] {$\i\ifnum\j<4\phantom{-}\fi$};
\foreach\i in {1}
  \draw (0.05,\i) -- (-0.05,\i) node [left,gray]  {$\i$};
	
\ifthenelse{\equal{#7}{1}}
{
	\colorlet{edgeColor}{green!75!black}
	\tikzset{vertexPartition/.style={draw, circle, inner sep=0pt, minimum size=.15pt, font=\tiny},
	dotVertex/.default = 0.5pt, 
	edgePartition/.style={draw, color=green!50!gray,line width=0.15mm},
	vertexOut/.style={inner sep=0pt, minimum size=.15pt},
	}	
  \draw[edgePartition]
	      (-3.8,-0.3) node[vertexPartition](V01){} 
		 -- (-2.71,-0.3) node[vertexPartition](V02){} 
		 -- (-2.1,-0.3) node[vertexPartition](V03){} 
		 -- (-1.6, 0.2) node[vertexPartition](V04){} 
		 -- (-0.8,-0.5) node[vertexPartition](V05){} 
		 -- (-0.2,-0.3) node[vertexPartition](V06){} 
		 -- (+0.6,+0.1) node[vertexPartition](V07){} 
		 -- (+1.6,-0.5) node[vertexPartition](V08){} 
		 -- (+2.6,+0.1) node[vertexPartition](V09){} 
		 -- (+2.5,+0.8) node[vertexPartition](V10){} 
		 -- (+2.2,+1.5) node[vertexPartition](V11){} 
		 -- (+1.6,+1.8) node[vertexPartition](V12){} 
		 -- (+0.3,+1.2) node[vertexPartition](V13){} 
		 -- (-0.6,+1.6) node[vertexPartition](V14){} 
		 -- (-1.2,+1.7) node[vertexPartition](V15){} 
		 -- (-1.9,+1.8) node[vertexPartition](V16){} 
		 -- (-2.2,+1.5) node[vertexPartition](V17){} 
		 -- (-3.1,+1.7) node[vertexPartition](V18){} 
		 -- (-3.6,+1.1) node[vertexPartition](V19){}
		 -- (V01);
  \draw[edgePartition] (V17) -- (V04);
	\draw[edgePartition] (V15) -- (V05);
	\draw[edgePartition] (V14) -- (V06);
	\draw[edgePartition] (V14) -- (V13) -- (V07);
	\draw[edgePartition] (V07) -- (V11);
	\draw[edgePartition] (V01) -- (-4.2,-0.75) node[vertexOut](VOut1){};
	\draw[edgePartition] (V02) -- (-2.52,-0.75) node[vertexOut]{};
	\draw[edgePartition] (V03) -- (-1.8,-0.75) node[vertexOut](VOut2){};
	\draw[edgePartition] (V05) -- (-1.0,-0.75) node[vertexOut]{};
	\draw[edgePartition] (V06) -- (+0.1,-0.75) node[vertexOut]{};
	\draw[edgePartition] (V08) -- (+1.78,-0.75) node[vertexOut]{};
	\draw[edgePartition] (V09) -- (+2.7,+0.1) node[vertexOut]{};
	\draw[edgePartition] (V10) -- (+2.7,+0.8) node[vertexOut]{};
	\draw[edgePartition] (V11) -- (+2.7,+1.64) node[vertexOut]{};
	\draw[edgePartition] (V12) -- (+1.57,+1.90) node[vertexOut]{};
	\draw[edgePartition] (V14) -- (-0.5,+1.90) node[vertexOut]{};
	\draw[edgePartition] (V15) -- (-1.18,+1.90) node[vertexOut]{};
	\draw[edgePartition] (V16) -- (-1.92,+1.90) node[vertexOut]{};
	\draw[edgePartition] (V18) -- (-3.12,+1.81) node[vertexOut]{};
	\draw[edgePartition] (V19) -- (-4.2,+1.3) node[vertexOut]{};
}{}

\def\arcStart{0.3*pi}
\def\arcEnd{1.4*pi}
\colorlet{pathColorWeak}{#3!#4}

\foreach\w in {#1,...,#2}
{
	\def\ww{(1+\w*\rw)}
	
	\ifthenelse{\equal{#5}{0}}
	{
		\ifcase\w	
			\colorlet{pathColor}{#3}
		\else			
			\colorlet{pathColor}{pathColorWeak}
		\fi
		
		\def\www{(abs(#7-1)+abs(\w))}
		\ifthenelse{\equal{#8}{1} \AND \equal{\w}{0}}{}
		{
			\draw[>->, >=stealth, line width=0.25mm, pathColor] plot[variable=\t,domain=\arcStart:\arcEnd,smooth,thick] ({paramFuncX(\t,\ww)},{paramFuncY(\t,\ww)});
		}
	}
	{
		\def\tStep{((\arcEnd-\arcStart)/#5)}
		\foreach\u in {1,...,#5}
		{
			\pgfmathparse{\u-abs(\w)>0 ? 1 : 0}
			\ifthenelse {\equal{\pgfmathresult}{1}}
			{
				\def\tStart{(\arcStart+(\u-1)*\tStep)}
				\def\tEnd{(\tStart+\tStep)}
				\def\tJoin{(\tStart+#6*\tStep)}
				\def\lWidthUnit{(0.15mm)}
				\def\lWidth{((#5+2-\u)*\lWidthUnit)}
				\def\lWidthNext{(\lWidth-\lWidthUnit)}
				
				\ifcase\u	
					\ifthenelse{\equal{\w}{0}}{\colorlet{pathColor}{#3}}{\colorlet{pathColor}{pathColorWeak}}
				\or				
					\ifthenelse{\equal{\w}{1}}{\colorlet{pathColor}{#3}}{\colorlet{pathColor}{pathColorWeak}}
				\or				
					\ifthenelse{\equal{\w}{0}}{\colorlet{pathColor}{#3}}{\colorlet{pathColor}{pathColorWeak}}
				\or				
					\ifthenelse{\equal{\w}{-1}}{\colorlet{pathColor}{#3}}{\colorlet{pathColor}{pathColorWeak}}
				\or				
					\ifthenelse{\equal{\w}{-2}}{\colorlet{pathColor}{#3}}{\colorlet{pathColor}{pathColorWeak}}
				\else			
					\colorlet{pathColor}{#3}
				\fi

				\draw[->, >=stealth, line width=\lWidth, pathColor] plot[variable=\t,domain=\tStart:\tJoin,smooth,thick] ({paramFuncX(\t,\ww)},{paramFuncY(\t,\ww)});
				\draw[line width=\lWidthNext, pathColorWeak] plot[variable=\t,domain=\tJoin:\tEnd,smooth,thick] ({paramFuncX(\t,\ww)},{paramFuncY(\t,\ww)});
				\foreach\uu in {-1,1}
				{
					\ifthenelse{\equal{\uu}{1}}
					{
						\ifcase\u	
							\colorlet{pathColorBranch}{pathColor}
						\or				
							\colorlet{pathColorBranch}{pathColor}
						\or				
							\colorlet{pathColorBranch}{pathColorWeak}
						\or				
							\colorlet{pathColorBranch}{pathColorWeak}
						\or				
							\colorlet{pathColorBranch}{pathColorWeak}
						\else			
							\colorlet{pathColorBranch}{pathColor}
						\fi
					}
					{
						\ifcase\u	
							\colorlet{pathColorBranch}{pathColorWeak}
						\or				
							\colorlet{pathColorBranch}{pathColorWeak}
						\or				
							\colorlet{pathColorBranch}{pathColor}
						\or				
							\colorlet{pathColorBranch}{pathColor}
						\or				
							\colorlet{pathColorBranch}{pathColor}
						\else			
							\colorlet{pathColorBranch}{pathColorWeak}
						\fi
					}

					\draw[bend right=\uu*15,line width=\lWidthNext, pathColorBranch] ({paramFuncX(\tJoin,\ww)},{paramFuncY(\tJoin,\ww)}) edge ({paramFuncX(\tJoin+(1-#6)*\tStep,\ww+\uu*\rw)},{paramFuncY(\tJoin+(1-#6)*\tStep,\ww+\uu*\rw)});
				}
			}{}
		}
	}
}
\ifthenelse{\equal{#2}{0}}
{
\foreach \i in {0,...,3}
{
	\def\t{(\i+1)*\arcStart}
  \node[circle, fill=#3, inner sep=1pt, label={[font=\small]45:{\text{t=}\i}}] at ({paramFuncX(\t,1)}, {paramFuncY(\t,1)}) {};
}
}{}

\ifthenelse{\equal{#8}{1}}
{
	\node at (-2.71,0.4) {\scalebox{3}{\twemoji{smirk_cat}}};
	\node at (-2.9,-0.6) {\scalebox{2}{\twemoji{smiley_cat}}};
	\node at (-3.2,-0.6) {\scalebox{1.5}{\twemoji{milk_glass}}};
	\node at (-2.35,-0.6) {\scalebox{2}{\twemoji{crying_cat_face}}};
	\node at (-2.05,-0.6) {\scalebox{1.5}{\twemoji{broccoli}}};
}{}
\end{tikzpicture}
}%

\newcommand{\classicalOneToOne}[4]
{
	\classicalManyToMany{#1}{#2}{red}{#3}{0}{1}{#4}{0}
}%

\newcommand{\bump}[2]
{
	\addplot[color=gray,left color=black,right color=black,middle color=white!95!black,samples=100,domain=#2-0.5:#2+0.5] {exp((-(x-#2)^2)*#1)*#1/pi};
}%

\newcommand{\evolves}[3]
{
	\addplot[-stealth, thick, color=blue] coordinates {(#1-#2,#3) (#1+#2,#3)};
}%

\newcommand{\wavefunction}[1]
{
\begin{tikzpicture}[yscale=1.25,xscale=2.0]
		\begin{axis}[view={0}{90},hide axis,
                 shader=interp,
                 mesh/color input=explicit mathparse]
        \addplot3[surf,unbounded coords=jump,
                  domain  =0:360,samples  =27,
                  domain y=0:360,samples y=41,
                  z buffer=sort,
                  point meta={symbolic={Hsb=#1==0?0:(\x+\y*\y/360-(\x+\y*\y/360>360?360:0)),#1==0?0:1,.5+cos(3*\x*\x/360-2*\y)/5}}]
            (x,y,1);
    \end{axis}
\end{tikzpicture}
}%

\newcommand{\coexistingdistribution}[1]
{
	\begin{tikzpicture}
	\begin{axis}[
			hide axis,
			y=0.5cm,
			x=8cm]
		
		\addplot[color=black,left color=gray!50!white,right color=gray!50!white,middle color=gray!50!white,samples=100,domain=-1.0:1.0] {exp((-(x-0.3)^2)*16)*16/pi+exp((-(x+0.3)^2)*8)*8/pi};
		\addplot[-stealth, color=black] coordinates {(-1.0,0) (1.0,0)};
		\node at (0.9,0.6) {$\mc{S}$};
			
		\ifthenelse{\equal{#1}{1}}
		{
		\addplot[color=blue,left color=blue,right color=blue,middle color=blue,samples=100,domain=0.34:0.4] {exp((-(x-0.37)^2)*10000)*9/pi};
		}{}

		\foreach\s in {-0.8,-0.6,-0.5,-0.38,-0.3,-0.22,-0.1,0,0.1,0.17,0.23,0.27,0.3,0.33,0.37,0.43,0.5,0.6,0.75}
			\addplot[-stealth,color=red!50!orange,very thick] coordinates {(\s,0) (\s,2)};
	\end{axis}
	\end{tikzpicture}
}%

\usepackage{authblk}

\newsavebox\affbox
\author{Cristi Stoica\ \orcidlink{0000-0002-2765-1562}}
\affil{Dept. of Theoretical Physics, NIPNE---HH, Bucharest, Romania.\\
Email: \textit{\color{cyan}\href{mailto:cristi.stoica@theory.nipne.ro}{cristi.stoica@theory.nipne.ro},  \href{mailto:holotronix@gmail.com}{holotronix@gmail.com}}}

\makeatletter
\newcommand*\@secondofsix[6]{#2}
\newcommand{\addtotitleformat}{%
  \@ifstar{\addtotitleformat@star}{\addtotitleformat@nostar}}
\newcommand\addtotitleformat@nostar[2]{%
  \PackageError{titlesec}{non starred form of \string\addtotitleformat\space not supported}{}}
\newcommand\addtotitleformat@star[2]{%
  \expandafter\expandafter\expandafter\expandafter
  \expandafter\expandafter\expandafter\def
  \expandafter\expandafter\expandafter\expandafter
  \expandafter\expandafter\expandafter\@currentsection@font
  \expandafter\expandafter\expandafter\expandafter
  \expandafter\expandafter\expandafter{%
    \expandafter\expandafter\expandafter\@secondofsix
       \csname ttlf@\expandafter\@gobble\string#1\endcsname}%
  \titleformat*{#1}{\@currentsection@font#2}%
}
\makeatother

\titlespacing\section{0pt}{20pt plus 6pt minus 4pt}{16pt plus 4pt minus 4pt}
\titlespacing\subsection{12pt}{12pt plus 6pt minus 4pt}{10pt plus 4pt minus 4pt}
\titlespacing\subsubsection{12pt}{12pt plus 6pt minus 4pt}{10pt plus 4pt minus 4pt}

\titleformat{\section}{\normalfont\fontsize{16}{24}\bfseries}{\thesection.}{1em}{}
\titleformat{\subsection}{\normalfont\fontsize{14}{20}\bfseries}{\thesubsection.}{1em}{}
\titleformat{\subsubsection}{\normalfont\fontsize{13}{18}\bfseries}{Substep \thesubsubsection.}{1em}{}

\titleformat{\author}{\normalfont\fontsize{14}{20}\bfseries}{\thesection}{1em}{}

\makeatletter
\renewcommand{\thesubsection}{Step \arabic{section}.\arabic{subsection}}
\makeatother



\definecolor{titcolor}{RGB}{0,90,255}
\addtotitleformat*{\section}{\Large\sffamily\color{titcolor}}
\addtotitleformat*{\subsection}{\large\sffamily\color{titcolor}}
\addtotitleformat*{\subsubsection}{\large\sffamily\color{titcolor}}

\title{\color{titcolor}\textbf{Classical Many-Worlds Interpretation}}

\date{\small\today} 

\begin{document}

\pagestyle{headings}	
\newpage
\setcounter{page}{1}
\renewcommand{\thepage}{\arabic{page}}

\maketitle
	
\begin{fabstract}
I present a simple baby-steps reconstruction of quantum mechanics as a fully classical theory. The most radical conceptual leap required is that there are many coexisting classical worlds, but even this is justified by the necessity of objective probabilities. These baby steps lead to a version of the many-worlds interpretation of quantum mechanics with built-in probabilities, built-in classicality at the macroscopic level, and an explanation of the complex numbers in quantum mechanics.
Despite its simplicity and minimalism of radical concepts, this is not a toy model, being equivalent with quantum field theory.
\end{fabstract}



\section{Introduction}
\label{s:intro}

Quantum mechanics seems to require huge jumps in our classical intuitions, shaped both by the history of physics, and mostly by the fact that we live in the macroscopic level of reality, which appears to us as classical.
In addition, there are important questions about the nature of reality, its independence of the observer, the nature of quantum probability, the emergence of the apparently classical world at the macroscopic level, the contradiction between the deterministic evolution described by the {\schrod} equation and the apparent wavefunction collapse, \etc.
The \emph{wavefunction collapse} was postulated to explain why, for a measurement with some possible results, the {\schrod} equation predicts that all of them are obtained, but we somehow see only one of them.

The \emph{many worlds interpretation} (MWI) appeared as the logical conclusion that all these results are obtained, but the {\schrod} equation itself shows that the wavefunction splits automatically into more components, more worlds, and the instance of the observer from each of these worlds is oblivious to the other worlds in which the other results occurred \citep{Everett1957RelativeStateFormulationOfQuantumMechanics,Everett1973TheTheoryOfTheUniversalWaveFunction,Wallace2012TheEmergentMultiverseQuantumTheoryEverettInterpretation,SEP-Vaidman2021MWI}. 

To many, MWI resolves all conceptual problems of quantum mechanics.
But outside the MWI community this solution is not accepted, and it is sometimes even fiercely rejected, and truth be told, even within the community there are strong disagreements \citep{Vaidman2024TheMWIofQMCurrentStatusAndRelationToOtherInterpretations}.
This may give the outsiders a feeling similar to that coming from religious disagreements,

\pagebreak

\begin{fquote}
``We are all atheists about most of the gods that humanity has ever believed in. Some of us just go one god further.''
\begin{flushright}
Richard Dawkins, ``The God Delusion'', 2006
\end{flushright}
\end{fquote}

A major point of divergence is how to obtain the probabilities we observe when repeating a quantum measurement with two possible outcomes with probabilities, for example, 1/3 and 2/3. Given that the wavefunction must split into two worlds, the probabilities seem to be cast in stone to 1/2 and 1/2.
Various solutions were proposed to obtain the probabilities, for example by world-counting, initiated by Everett himself \citep{Everett1957RelativeStateFormulationOfQuantumMechanics} and refined in numerous articles like \citep{Saunders2021BranchCountingInTheEverettInterpretationOfQuantumMechanics,Saunders2024FiniteFrequentismExplainsQuantumProbability}, or from ignorance after the measurement \citep{Vaidman1998OnSchizophrenicExperiencesOfTheNeutronOrWhyBelieveMWI}, or from symmetry \citep{Vaidman2012ProbabilityInMWI}, or from the idea that decision theory should force us to accept the right probabilities as true \citep{Deutsch1999QuantumTheoryOfProbabilityAndDecision,Wallace2002QuantumProbabilitiesAndDecisionRevisited}, and so on. For a review see \citep{Vaidman2020DerivationsOfTheBornRule}.

Here I introduce a conceptually simple, step-by-step reconstruction of quantum mechanics, that gives probabilities as simple ignorance of the observers about their existence in one or another of many classical worlds.
The steps are really simple and each of them leads to a still classical theory, so that even someone who lived before the discovery of quantum mechanics should have little trouble accepting their possibility.
In this article, I will try to make it very accessible, but more technical details can be found in \citep{Stoica2022BornRuleQuantumProbabilityAsClassicalProbability,Stoica2022BackgroundFreedomLeadsToManyWorldsLocalBeablesProbabilities,Stoica2023TheRelationWavefunction3DSpaceMWILocalBeablesProbabilities}.
Since the worlds are classical, this entails an almost evaporation into thin air of other problems, like why the world appears to us classical at the macroscopic level, or why complex numbers appear in quantum mechanics. 

The reader may find this solution too simple, but I will show that it has all the features of quantum field theory.

\section{From classical to quantum probabilities}
\label{s:steps}

To make sense of probabilities in MWI, we will start with classical physics. 

\begin{ffig}
\begin{center}
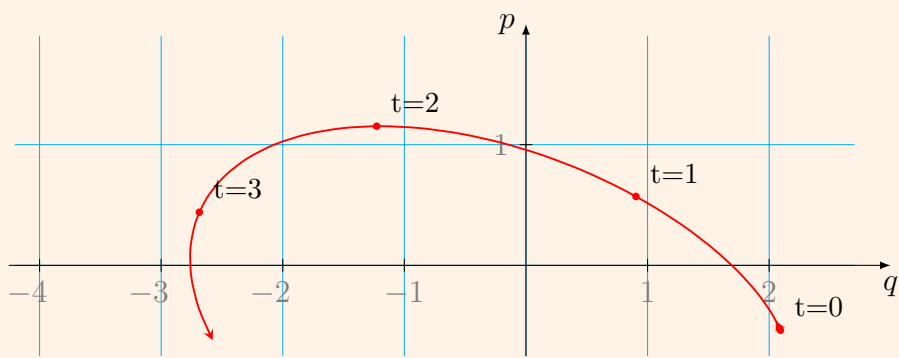

\classicalOneToOne{0}{0}{0}{0}
\captionof{figure}{Single deterministic classical history of the world. The trajectory represents a succession of states of the world at different times.}
\label{fig:classical:deterministic:single:cert}
\end{center}
\end{ffig}

In classical physics, each possible state can be uniquely identified by the values of a number of physical properties of the system in that state.
That is, these properties are chosen so that any two distinct states differ by the values of at least one of these properties.
For example, a list of properties characterizing a system of $n$ point-particles can consist of the $x,y,z$ coordinates and the $x,y,z$ components of the momenta of each particle.
Then, each state can be represented uniquely as a point in a \emph{state space}, so that its coordinates in the state space are the values of these properties.
Think of this as similar to the latitude and the longitude that uniquely identify a place on the surface of the Earth, or, if you prefer, similar to the address of each home.

It should be clear by now that the value of any other property of the physical world is determined by the properties used to parametrize its states, so it is a function of these properties.

Classical physics is deterministic, so the initial state of the world at the time $t_0$ determines its state at all other times, for example at subsequent times $t_1,t_2,\ldots$, as represented in Fig. \ref{fig:classical:deterministic:single:cert}.

In the following, by ``deterministic'' I will understand that both the future and the past can be uniquely determined by the present state and the evolution law. By ``indeterministic'' I will understand the opposite.

The state of the world is known only by Laplace's daemon. To us mortals, only an extremely coarse approximation of the state is accessible, what we see at the macroscopic level.
We call these approximations \emph{macrostates}, since they are not states (points in the state space), but extended regions, as represented in Fig. \ref{fig:classical:deterministic:single:cert:macro}. To avoid any confusion between states and macrostates, sometimes the states are called \emph{microstates}. As in the case of the microstates, each macrostate can be identified uniquely by a number of properties, so that these properties have equal values for all of the states in the same region. If you think of a microstate as similar to the address of somebody, you can think of the values of the macroscopic properties as similar to the zip code, because different homes can have the same zip code but different addresses.
Let's call these properties, and all properties that can be derived from them (that is, they are functions of them), \emph{macroproperties}.

\begin{ffig}
\begin{center}
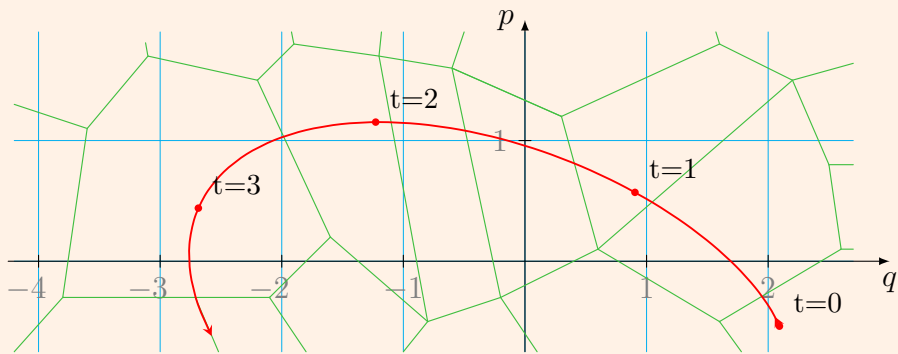

\classicalOneToOne{0}{0}{0}{1}
\captionof{figure}{Macrostates are extended regions of the state space, and represent the limited knowledge available to us about the microstate. During its evolution, the world goes from macrostate to macrostate.}
\label{fig:classical:deterministic:single:cert:macro}
\end{center}
\end{ffig}

You can try to make the most detailed picture of the world, using as many photos and videos and words and sounds as you want, you'll never be able to describe its microstate.
In fact, not even its macrostate, because even about this you can only have limited information. But in the following I will assume that we can know any macroproperty, at least in principle.

Nevertheless, you can also know microproperties by using a microscope or any other instrument that allows you to make a microproperty affect a macroproperty.
For example, if you measure the size of a microscopic object with the help of the microscope, what you do in fact is to obtain a macroscopic image of that object, and to measure that image, and deduce from this the size of the microscopic object.
Something like this happens in the case of all classical measurements.

What you do when you measure a microproperty is to arrange an experiment so that the microproperty affects a macroproperty in a predictable way.
But microproperties affect macroproperties all the time, because macroproperties are functions of the microproperties. Since the microproperties are poorly known to us, they often appear to us as random events.
For example, if we flip a coin, we don't know what the result will be.
But for Laplace's demon, who knows the microstate, knowing this is a piece of cake.

The fact that probabilities occur in a classical world is uncontroversial, so let us start with this.
Then, step-by-step, we will introduce probabilities in MWI in a classical style.

\subsection{Not knowing the real state of the world}
\label{s:steps:classical:deterministic:single:prob}

If we flip a fair coin in a classical deterministic world, the probability of getting heads is about the same as that of getting tails. But whatever result we will obtain, this is already written in the properties of the microstate of the world before flipping the coin, and in fact in the very initial conditions of the world. The only reason the result is unknown to us before looking at the coin is that we don't know the microstate of the world in full detail. We can only know the approximate values of some of its macroproperties. The best we can do is to consider various possible states of the world, as in Fig. \ref{fig:classical:deterministic:single:prob}, and to estimate what is the probability that the world is in one or another of these states.
The actual state is one of them, only one. We don't know which, but we have a guess of the probability for each case.

\begin{ffig}
\begin{center}
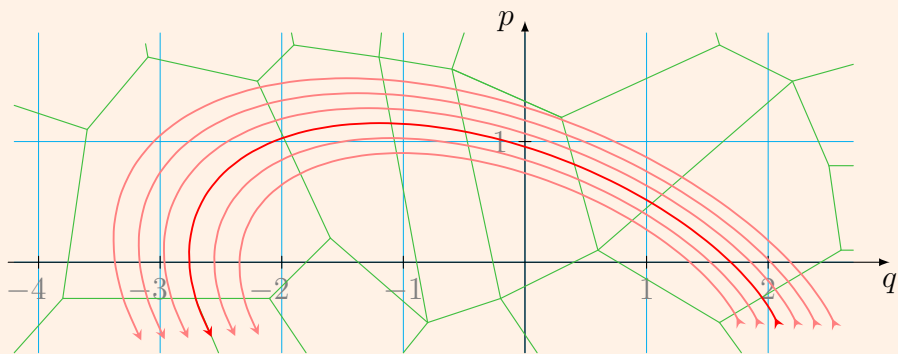

\classicalOneToOne{-2}{3}{50}{1}
\captionof{figure}{Possible classical classical histories of a deterministic world. Only one of these histories is the real one, but we just don't know which.}
\label{fig:classical:deterministic:single:prob}
\end{center}
\end{ffig}

Therefore, in a classical world, the best we can do is to make guesses about the state of the world, about the values of its properties.

\begin{frem}
Probabilities are quantitative expressions of our guesses.
\end{frem}

We can make such guesses even if the evolution is indeterministic, that is, as shown in Fig. \ref{fig:classical:nondeterministic:coexisting:prob}, the initial state at $t_0$ doesn't uniquely determine the future state at $t_1,t_2,\ldots$.

\begin{ffig}
\begin{center}
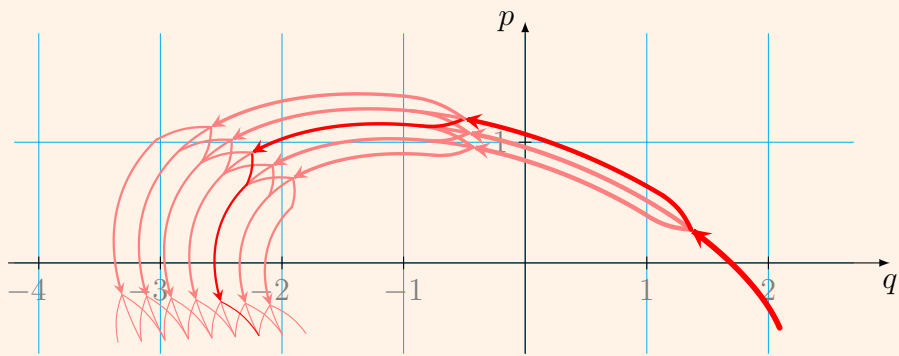

\classicalManyToMany{-3}{3}{red}{50}{4}{0.8}{0}{0}
\captionof{figure}{Possible histories of a single indeterministic classical world.}
\label{fig:classical:nondeterministic:single:prob}
\end{center}
\end{ffig}

In this case, when estimating the probabilities, we have to take into account the probabilities that are already built in the indeterministic theory.

But even if we can conceive probabilities in classical worlds, even if we can conceive indeterministic classical worlds, these are not quantum worlds.
At least not yet, not in this form.

\subsection{Not knowing in which one of more worlds you are}
\label{s:steps:classical:deterministic:coexisting:prob}

It is important to highlight that our guesses are based on our incomplete knowledge of the microstate of the world.
Therefore, 

\begin{frem}
The probabilities that we assign are subjective, and not an objective feature of the world.
\end{frem}

If we think seriously about classical probabilities, we may realize that the idea of an objective probability that the world is in this or that state doesn't make much sense.
The reason is that \emph{there is a unique history}. It makes no sense to assign probabilities to what other possible histories could have happened. It's not that we don't know their probabilities, the point is that they didn't happen, only one world happened. We can't even know much in what other possible states the world \emph{could} have been, if it never was and will never be in any of these states. We think that they should correspond to all possible combinations of values of the parameters, and so we postulate the existence of a state space of all possible worlds, and then we make guesses about the probability that the world is in one or another state. And even if our guess turns out to work in practice, since only one of the histories contained in the state space is the real history, its probability is one, and as long as other possible histories never occurred, their probability is simply zero.

We tend to assume the existence of an objective probability that the world is in one state or another so that we can use it to make some predictions. We can predict, for example, that the chance that the result of tossing a fair die is \raisebox{-1pt}{\scalebox{1.25}{\epsdice{3}}} is $1/6$.
Or we can make use of probabilities in statistical mechanics, for example to describe the entropy of a system. The entropy is physical, so if its cause is due to statistics, probabilities must be objective, right?
But our notions of probabilities are idealizations, objectifications of subjective probabilities, so how do we get objective probabilities that have physical effects like entropy?

A way to get probabilities is to flip a coin repeatedly in the same world, and count how many times we've got tails and how many times we've got heads. But no matter how fair the coin is, it would be extremely unlikely to get heads and tails the same number of times. The frequency with which each result occurs surely allows us to extract \emph{subjective} probabilities, and these probabilities seem to converge, in many but not all cases and for a large number of repetitions, to a limit that we may hope to be objective. But this convergence itself is subject to probabilities, and for each possible sequence in which it happens, there are infinitely many other sequences in which it doesn't happen. This is what the theory of probabilities itself says. Therefore, the theory of probabilities itself informs us that the frequency of occurrence of heads can't be used to define a precise notion of objective probability.

Another way to get probabilities is to assume that all possible outcomes exist in an abstract sense, as possibilities, or as propensities, but not physically. That is, we assume them to be real just from the point of view of alternative histories, but unreal otherwise.

\begin{fquest}
My simple mind can't comprehend this, to me either they are real, or they are not real at all. I don't say this is impossible, only that I couldn't make sense of such a claim.
It seems that such probabilities can't be objective, but only subjective.
\end{fquest}

However, we can imagine situations in which probabilities are objective.
Consider a fictional situation in which you are destroyed and, instantaneously, you are reconstructed with perfect accuracy, down to the level of particles composing your body, in two distinct copies. Suppose that each of these copies will be in one of two isolated rooms, labeled by the numbers $1$ and $2$.
Then, it makes sense for you to think, after this duplication takes place, that the probability that you are in room number $1$ is $1/2$, and that this is an objective probability.
There is still a single world, but there are two instances of you in this world, so now you can talk about objective probabilities even if there is only one world. These are \emph{self-location probabilities}.

\begin{frem}
Self-location probabilities are also subjective, because they come from your ignorance of knowing which of the two copies you are. But at the same time they are also objective, because they are not about what could have happened but didn't happen, they are about what really happened.
\end{frem}

In fact, we don't even need to imagine such fictional situations, because it happens to us every time.
There are multiple instances of yourself, just not at the same time.
Suppose you wake up after sleeping. If you are like most people, you don't know what time it is. It could be \texttt{6:37}, it could be \texttt{7:23}, you just don't know without looking at a clock. Another way to put it is that you don't know if you are the \texttt{6:37} instance of yourself, or the \texttt{7:23} version, or any other version of yourself within a reasonable waking-up time interval.
These probabilities are still of self-location, but in time.

But what if you fell asleep on a train or a plane? Then, upon waking up, you can ask yourself not only what time it is, but also where along the route you are. This is self-location in space.

Even the extraction of probabilities from the frequencies of the results of repeated experiments is about self-location: just like trying to guess the time or place is about self-location in time and respectively space, estimating the probability of getting tails from repeatedly flipping a coin is about self-locating yourself among the events of flipping the coin.
And if more people run independent repetitions in different places in the world, and then combine the results, you can see the frequentist probabilities as about self-location among the events of flipping the coin.

Similarly, if more distinct worlds were to exist and evolve in parallel as in Fig. \ref{fig:classical:deterministic:coexisting:prob}, it would make sense to consider probabilities as objective.
This time they are not about what could have happened although it never happened, but about something that happened in multiple distinct instances. The dual subjective-objective nature of self-location probabilities is due to our ignorance of the world in which we are, on the background of more worlds being real.

\begin{ffig}
\begin{center}
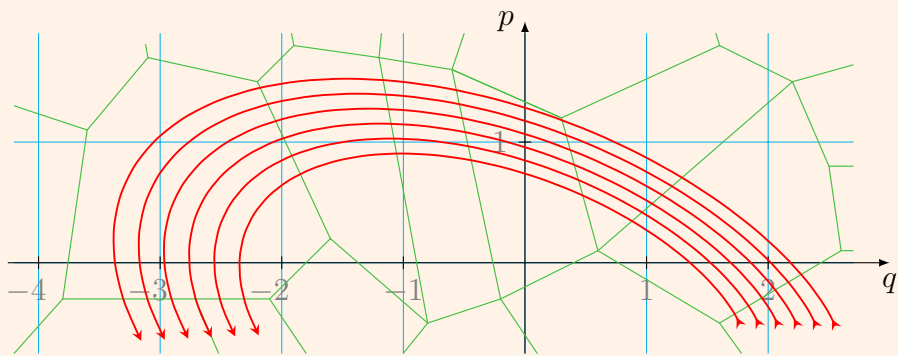

\classicalOneToOne{-2}{3}{100}{1}
\captionof{figure}{Coexisting deterministic classical worlds. All worlds exist, we just don't know in which we are.}
\label{fig:classical:deterministic:coexisting:prob}
\end{center}
\end{ffig}

Self-location probabilities as introduced here still refer to classical worlds. We just allow for more classical worlds to exist simultaneously.
And this offers a way to make sense of objective probabilities that a classical world is in one or another possible state, an objective ground to our subjective estimates of these probabilities.

\begin{frem}
Coexisting classical worlds give objectivity to probabilities and entropy.
\end{frem}

And indeed, if the world could happen once, why wouldn't it happen more times, in different versions?
There is of course the question of empirical evidence: if each world evolves independently, there is no way to know if the other worlds exist. Why invent them just to give an objective basis to the probabilities that the world could have been different? This seems too extravagant! We will return to this question later.

\subsection{Identity-changing classical worlds}
\label{s:steps:identity}

In this article we focus on self-location probabilities about being in one or another of more possible worlds.
For a deterministic world, since the state of the world at a given time determines entirely its future history, the self-location probability of being in one history or another is the same as the self-location probability of being in one world or another.
The evolution of a classical world, as described in classical physics, preserves the identity of the world, in the sense that distinct histories don't intersect. There is a continuity of the identity of a world.

A major difference between quantum and classical physics is that whatever you would do in a world described by classical physics, it will not contradict what you knew about the state of the world. For example, if you learn the result of flipping a coin, this is perfectly consistent with whatever other things you previously knew about the world.

This is no longer true in a quantum world, where you can measure the position of a particle twice and get different results. The reason is that the first position measurement makes the momentum indeterminate, so where the particle will be shortly after that becomes indeterminate as well.
It seems that whenever you extract new information about the world, some of the previous information becomes invalidated.
And this is not simply because any measurement interacts with the observed system -- it seems that the evolution equation itself is violated. If the {\schrod} equation predicts that the observed system will be in a particular state, measuring it may reveal that it is in a different state. Taking into account the interactions between the measurement device and the observed system doesn't solve this problem, the system seems to jump from the state in which was supposed to be according to the {\schrod} equation in one of the states in which can be expected to be according to the property that we measure, which in general is different.

This means that in a quantum world there are discontinuities in the evolution, or so it seems. The world itself is unable to maintain its identity across time.

However, we can imagine classical laws according to which the evolution is indeterministic, as in Fig. \ref{fig:classical:nondeterministic:coexisting:prob}. 

\begin{ffig}
\begin{center}
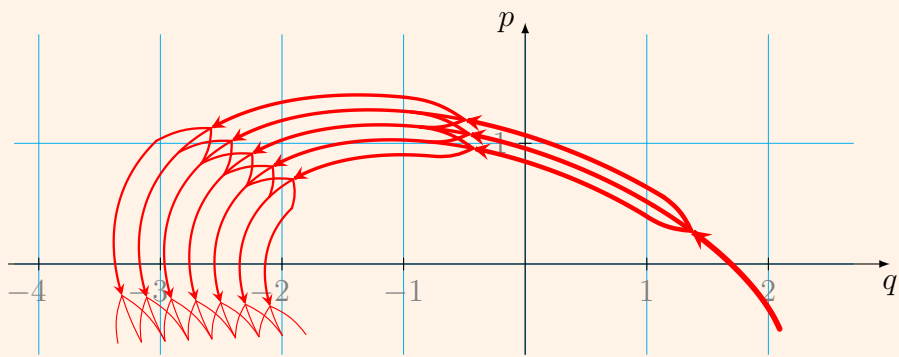

\classicalManyToMany{-3}{3}{red}{100}{4}{0.8}{0}{0}
\captionof{figure}{Coexisting indeterministic classical worlds. The evolution is indeterministic for a single world, but deterministic for all taken together.}
\label{fig:classical:nondeterministic:coexisting:prob}
\end{center}
\end{ffig}

\begin{frem}
The evolution is indeterministic for a single world, but deterministic for all taken together.
\end{frem}

In this case the identity of a world is not preserved, there is no continuity of the identity, the identity changes suddenly, as in Fig. \ref{fig:classical:nondeterministic:coexisting:prob}.
But this doesn't affect the fact that it makes sense to talk about self-location probability at a given time. 

At a given time the continuity of identity is irrelevant, it makes sense to talk about self-location probability whether or not the identity is continuous or discontinuous during the previous or subsequent evolution of the world. Assuming that more worlds coexist simultaneously, self-location about in which world you are is independent of the past history of the system, to which we no longer have access anyway except for what is recorded in the present state. Self-location probabilities are instantaneous, as in Fig. \ref{fig:classical:many:coexisting}.

\begin{ffig}
\begin{center}
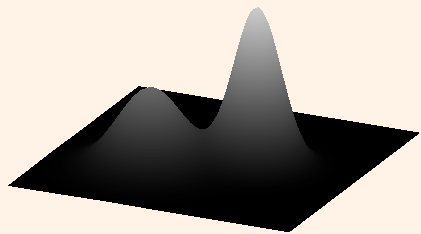

\begin{tikzpicture}
\begin{axis}[hide axis]
    \addplot3 [
        surf,
        shader=interp,
				opacity = 1,
				colormap name=mycolormap,
        samples=70,
        domain=-2.25:4.25,
    ] {exp(-x^2-y^2)/pi+exp((-(x-2)^2-(y-1)^2)*2)*2/pi};
\end{axis}
\end{tikzpicture}
\captionof{figure}{Probability distribution of coexisting worlds at a given time.}
\label{fig:classical:many:coexisting}
\end{center}
\end{ffig}

The difference between deterministic and indeterministic worlds is that indeterministic evolution allows the identities of the worlds to change. Thus, the probability distribution doesn't simply evolve while maintaining its shape over time, but it can change, for example it can spread, as in Fig. \ref{fig:classical:many:identity}.

\begin{ffig}
\begin{center}
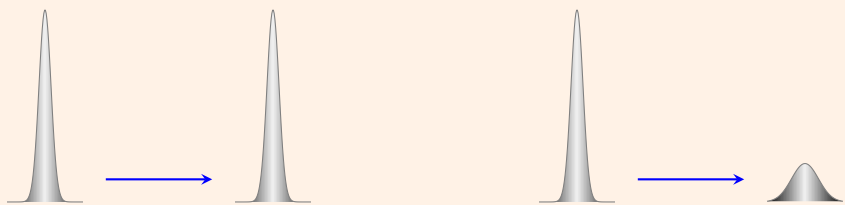

\begin{tikzpicture}
\begin{axis}[
		hide axis,
		y=0.1cm,
    x=1.0cm]
\bump{80}{0}
\evolves{1.5}{0.7}{3}
\bump{80}{3}
\bump{80}{7}
\evolves{8.5}{0.7}{3}
\bump{16}{10}
\end{axis}
\end{tikzpicture}
\captionof{figure}{\textbf{Left:} Classical deterministic evolution preserves the identity of the worlds, and therefore the localization of the probability distribution.
\textbf{Right:} Classical indeterministic evolution may spread the probability distribution.}
\label{fig:classical:many:identity}
\end{center}
\end{ffig}

And this brings us one step closer to the quantum worlds, in which, as a result of quantum measurements, a system can jump into a state that doesn't follow deterministically from the {\schrod} equation.
And therefore this is true of the entire world containing that system.

\subsection{Counting classical worlds}
\label{s:steps:counting}

A cat is waiting to be fed, and will get either milk or broccoli.
Fig. \ref{fig:schrod:cat:counting} depicts the macrostate in which the cat is waiting to be fed \emoji{smirk_cat}, and two subsequent macrostates in which it gets milk \emoji{milk_glass}\emoji{smiley_cat} and respectively broccoli \emoji{crying_cat_face}\emoji{broccoli}.

\begin{ffig}
\begin{center}
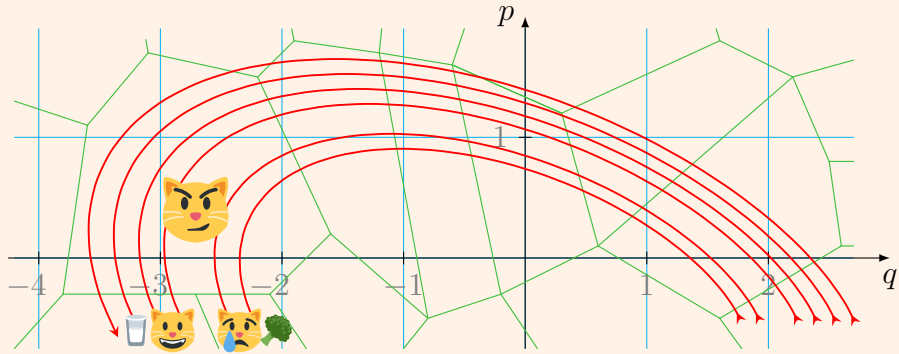

\classicalManyToMany{-2}{4}{red}{100}{0}{1}{1}{1}
\captionof{figure}{The probability that the cat gets milk can be obtained by counting.}
\label{fig:schrod:cat:counting}
\end{center}
\end{ffig}

Suppose there are two worlds in which the cat gets broccoli, and four worlds in which it gets milk.
Then, the probability that the cat gets milk is twice the probability that she gets broccoli.
In this case, we simply applied the classical formula for probabilities,

\begin{equation}
\label{eq:prob-classic}
p\(\text{the cat gets milk}\)=\frac{\text{number of worlds in which the cat gets milk}}{\text{total number of worlds}}=\frac{4}{6}=\frac{2}{3}.
\end{equation}

We can use a simple counting rule, as long as the number of worlds is finite.
But, if at some time there is a finite number of worlds, what happens if the probabilities change to another value, continuously?
If, as in \ref{s:steps:identity}, the classical worlds don't preserve their identities, they mix with one another. And this happens continuously, because the evolution equation is continuous.
So it would be impossible to have all the time probabilities that can be expressed as ratios as in equation \eqref{eq:prob-classic}.

\begin{fquest}
What would possibly be the meaning of fractional coexisting worlds of the form
\begin{equation}
\label{eq:coexisting-fractional}
\frac{2}{3}\text{``first classical world''}+\frac{1}{3}\text{``second classical world''}?
\end{equation}
\end{fquest}

This can't work if we assume that there is always an integer number of worlds. But maybe there are four copies of the first world and two copies of the second world. But what would it mean for a classical world to exist twice? 
The example with the person that is cloned in two copies, which was used in \ref{s:steps:classical:deterministic:coexisting:prob} to exemplify self-location probability, doesn't seem to support the idea that we can be ``more'' one of the clones than the other, the probability can only be the same.
Similarly, if only two classical worlds exist and contain instances of you, what could make the self-location probability be anything but 50/50?
Maybe it's possible, but fortunately we don't have to deal with this conceptual problem, because

\begin{frem}
If there are infinitely many classical worlds, a continuum of classical worlds, the coexisting worlds can equally exist but be distributed continuously and unevenly, as in Fig. \ref{fig:classical:continuous-counting-a}.
\end{frem}

\pagebreak

\begin{ffig}
\begin{center}
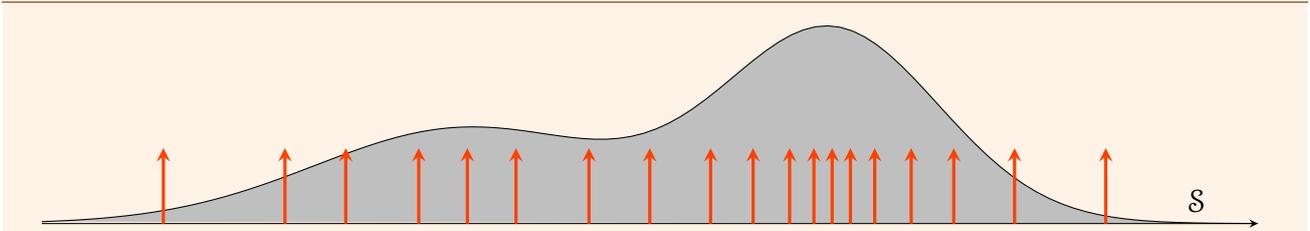

\coexistingdistribution{0}
\captionof{figure}{Equally existing coexisting classical worlds can be distributed more or less densely, if the state space $\mc{S}$ forms a continuum. The \textcolor{red!50!orange}{red arrows} represent a finite selection of the classical worlds, and the gray profile represents the continuous distribution.}
\label{fig:classical:continuous-counting-a}
\end{center}
\end{ffig}

Fig. \ref{fig:classical:continuous-counting-b} illustrates the continuous limit of a probability distribution of classical worlds, this time representing the state space as the plane.

\begin{ffig}
\begin{center}
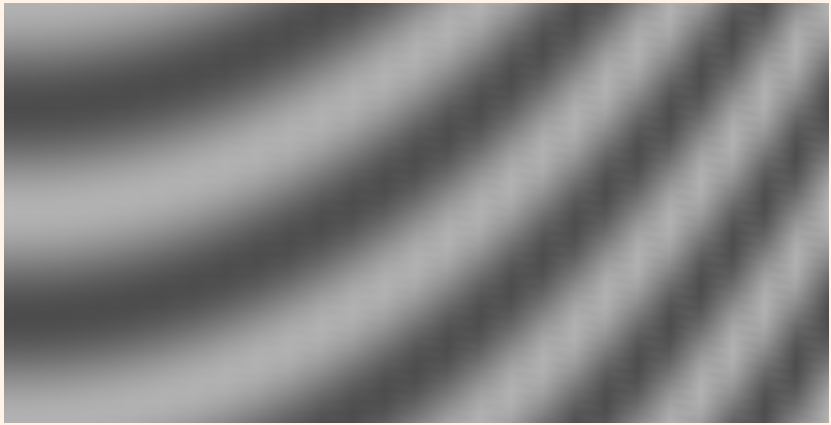

\wavefunction{0}
\captionof{figure}{A continuum of equally existing coexisting classical worlds, distributed unevenly (denser in the brighter regions and less dense in the darker regions).}
\label{fig:classical:continuous-counting-b}
\end{center}
\end{ffig}

And this makes sense, since in our reality there seem to be a continuum of possible classical states, and the evolution is also continuous.

And this brought us closer to an understanding of the wavefunction.

\subsection{Real wavefunction}
\label{s:steps:wavefunction-real}

The probability distributions are such that the total probability is one.
That is, the integral of the probability distribution $p(s,t)$ over all points $s$ (representing classical states) in the state space $\mc{S}$, at any time $t$, is one,
\begin{equation}
\label{eq:class-prob}\int_{\mc{S}} p(s,t)d\mu(s)=1.
\end{equation}

The integration is with respect to the measure $\mu(s)$ on the state space.
We are still talking about classical worlds, and the probability distribution itself is not yet a wavefunction as in quantum mechanics.
But if we collect the square roots of its values at all points in the state space, we obtain, for any $t$, a wavefunction $\psi(t):\mc{S}\to\R$, defined by
\begin{equation}
\label{eq:psi-real}
\psi(s,t):=\sqrt{p(s,t)}.
\end{equation}
Then, $\psi$ is a wave propagating not in the three-dimensional space, but in the state space.
For example, the wavefunction from Fig. \ref{fig:classical:wavefunction} gives the probability distribution from Fig. \ref{fig:classical:many:coexisting}.

\begin{ffig}
\begin{center}
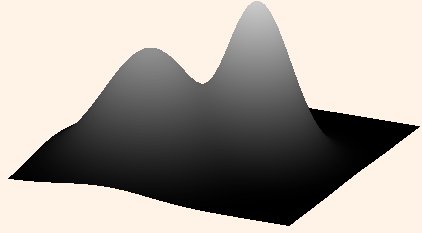

\begin{tikzpicture}
\begin{axis}[hide axis]
    \addplot3 [
        surf,
        shader=interp,
				opacity = 1,
				colormap name=mycolormap,
        samples=70,
        domain=-2.25:4.25,
    ] {sqrt(exp(-x^2-y^2)/pi+exp((-(x-2)^2-(y-1)^2)*2)*2/pi)};
\end{axis}
\end{tikzpicture}
\captionof{figure}{The wavefunction giving the probability distribution from Fig. \ref{fig:classical:many:coexisting}.}
\label{fig:classical:wavefunction}
\end{center}
\end{ffig}

This is like in quantum mechanics.
Any two such wavefunctions can be added, and the result is \emph{normalized}, resulting in a wavefunction
\begin{equation}
\label{eq:psi-add}
\psi(s,t):=\psi_1(s,t)+\psi_2(s,t) \text{\ \ \ \ (normalized)}
\end{equation}
where ``normalized'' means that the function $\psi_1(s,t)+\psi_2(s,t)$ is divided by a number so that the integral of $\psi^2(s,t)$ is one. 
Then $\psi^2(s,t)$ represents the probability distribution of the classical worlds $s$ at the time $t$.
Similarly, we can define linear combinations $a\psi_1+b\psi_2$ (normalized), where $a$ and $b$ are non-negative real numbers.

\begin{fquest}
But the wavefunction defined above has only real non-negative values, while in quantum mechanics the wavefunction is complex.
\end{fquest}

\subsection{Complex wavefunction and quantum fields}
\label{s:steps:wavefunction-complex}

In \emph{quantum field theory}, the wavefunction is complex only for charged particles and gauge fields, as we see when we quantize the field equations, see for example \citep{ItzyksonZuber2005QFT,Srednicki2007QFT,Zee2010QFTNutshell}.

This is true in particular in the \emph{wavefunctional} formulation of quantum field theory \citep{Hatfield2018QuantumFieldTheoryOfPointParticlesAndStrings}, which was shown to be equivalent with the other formulations.
In this formulation, which is obtained naturally by applying the method used by {\schrod} to quantize systems of point-particles, we start with a set of classical field configurations instead of classical point-particle configurations. Each of these field configurations is represented by a vector in a sufficiently high-dimensional vector space, so that any two distinct classical field configurations correspond to two orthogonal vectors.
The set of orthogonal vectors representing classical field configurations form a basis of the vector space, which is the state space for quantum mechanics.
Then, the wavefunction becomes a \emph{wavefunctional} dependent on the classical field configuration $s$, usually denoted in the form $\psi[s]$.
The word ``functional'' means a function that depends on other functions or fields, in this case classical field configurations.
But it's still a function, so there is no harm in continuing to call $\psi$ a wavefunction and write $\psi(s)$, as long as we remember that the classical state $s$ is a field configuration.

Classical electromagnetism is properly understood as a \emph{gauge theory} \citep{Georgi1982LieAlgebrasInParticlePhysics,BaezMunian1994GaugeFieldsKnotsGravity,Nakahara2003GeometryTopologyPhysics}.
Electromagnetism is associated to the existence of an internal symmetry space which is just a circle associated to each point of space, and the gauge is an angle identifying a point of the circle.
A \emph{local gauge} is a choice of an angle, identifying a point on the circle at each point of space.
A \emph{global gauge transformation} is a rotation of all local angles at once with the same angle $\varphi$ in the internal space.
For example, charged fields are multiplied by a global phase factor, which is a complex number $e^{i\varphi}=cos\varphi+i\sin\varphi$, where $\varphi$ is real, and $i$ represents a $90^\circ$ rotation in the internal space. Global gauge transformations have no observable physical effects. Then, an unobservable global phase can be associated to each classical field configuration for electrically charged fields and electromagnetic gauge fields.
Because it's unobservable, when representing a classical field configuration $s$ by a vector, let's denote it by $\ket{s}$, the phase is left aside, that is, $\ket{e^{i\varphi}[s]}$, where $e^{i\varphi}[s]$ is the classical field obtained by the global gauge transformation given by rotating $s$ with the angle $\varphi$ in the internal space, is taken to be the same vector $\ket{s}$.

On the other hand, the quantum wavefunctionals are complex, that is, $\psi(s,t)=r(s,t) e^{i\varphi(s,t)}$ is a complex functional, where $r(s,t)$ and $\varphi(s,t)$ are real functionals.
The phase $\varphi(s,t)$ is also unobservable, but phase differences reveals themselves in superpositions and interference, and in the Aharonov-Bohm effect \citep{Nakahara2003GeometryTopologyPhysics}.

So on one hand a gauge phase $\varphi$ is left aside when quantizing the classical fields, and an apparently unrelated phase $\varphi$ is added back to make the wavefunctional complex, as required by superposition and interference effects.
So why not skip the removal of the gauge phase and the inclusion of the complex phase, and just take them as being the same?
Then we will have
\begin{equation}
\label{eq:gauge-phase}
\ket{e^{i\varphi}[s]}=e^{i\varphi}\ket{[s]}.
\end{equation}

What I described so far is standard quantum field theory in the wavefunctional formulation, with the amendment that the phase of the complex wavefunctional is just the global gauge of the individual classical field configurations.
But is there something similar in a classical theory as introduced in the previous steps?
Indeed, there is. 
If more classical worlds coexist and the evolution doesn't preserve their identities, when two or more worlds evolve into the same state, as in Fig. \ref{fig:classical:nondeterministic:coexisting:prob}, the global gauge of each of these worlds has to be taken into account to obtain the world into which they combine, because these gauges contribute essentially to the result. In \ref{s:steps:wavefunction-real} we introduced classical dynamics of more coexisting worlds, and a real wavefunction that propagates on the state space. But if the global gauge of each individual world is taken into account, we obtain a complex wavefunction propagating on the state space.

\begin{ffig}
\begin{center}
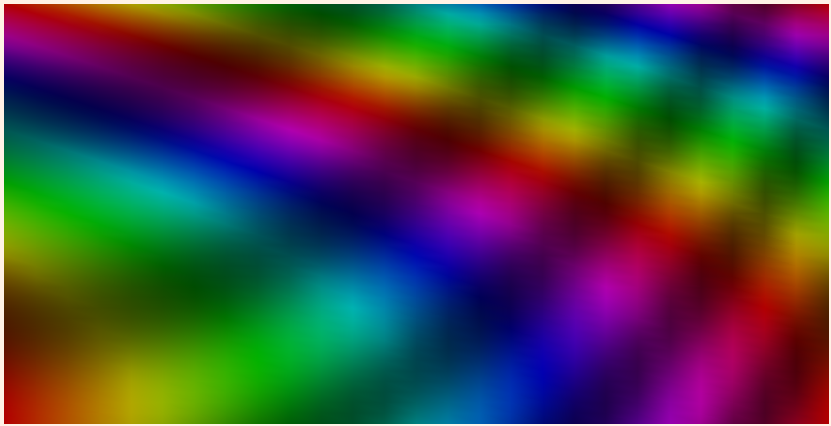

\wavefunction{1}
\captionof{figure}{Complex wavefunction from classical worlds, $\psi(s)=r(s) e^{i\varphi(s)}$. The color's hue represents the phase $\varphi(s)$, and its brightness represents the amplitude $r(s)$, where $s$ is from the classical field configuration space.}
\label{fig:classical:complex-wavefunction}
\end{center}
\end{ffig}

So now we see that 

\begin{frem}
The wavefunctional quantization as described here is just classical physics with coexisting worlds distributed according to the probability distribution $p(s,t)=r^2(s,t)$ as in \ref{s:steps:wavefunction-real}, with an identity-changing evolution as in \ref{s:steps:identity}, and whose evolution also involves the global gauges of the individual classical worlds.
\end{frem}

Therefore, quantum field theory is a particular case of a classical theory of this type.

\begin{frem}
This gives a ground for the probabilities, as arising just as in classical physics, but with coexisting worlds instead of a single world whose state is unknown.
\end{frem}

\section{Open future for Schr\"odinger's cat}
\label{s:indetermined}

In our quantum universe, if we measure the spin of a particle along the axis $z$, if the particle is prepared so that it's spin is oriented along the axis $x$, there is a fifty-fifty chance to get the spin up.
Then, presumably, if the spin is up the pointer of the measuring device indicates up, and if the spin is down, the pointer indicates down, as in Table \ref{table:spin-measurement}.

\begin{ftab}
\centering
\begin{tabular}{c|cc} 
					& The spin is up & The spin is down \\
 \hline
 \begin{tabular}{@{}c@{}}The pointer shows\\spin ``up''\end{tabular} & {\LARGE\textcolor{green!65!black}\cmark} & {\LARGE\textcolor{red}\xmark} \\
 \begin{tabular}{@{}c@{}}The pointer shows\\spin ``down''\end{tabular} & {\LARGE\textcolor{red}\xmark} & {\LARGE\textcolor{green!65!black}\cmark} \\ 
\end{tabular}
\captionof{table}{Quantum measurements copy (or even ``create'' if the property's value was undefined) information about the state of the observed system in the state of the pointer.}
\label{table:spin-measurement}
\end{ftab}

Following {\schrod}, let's connect the measuring device to a machine that feeds a cat, so that if the spin turns out to be up, the cat gets milk, and if the spin turns out to be down, the cat gets broccoli. Why do we say that the {\schrod} equation predicts that the cat receives only milk but also only broccoli?
The reason is that, according to {\schrod}'s equation, the wavefunction of the world before the measurement evolves into the sum of the wavefunction of a world in which the spin was up and the cat received milk, and the wavefunction of a world in which the spin was down and the cat received broccoli.
The measurement problem asks

\begin{fquest}
If {\schrod}'s equation predicts that both these outcomes should happen in a superposition, why only one of them seems to happen?
\end{fquest}

Let's see how the classical theory of coexisting worlds from \ref{s:steps:wavefunction-complex} explains this.

The initial macrostate is so that
\begin{itemize}
	\item
there is a hungry cat \emoji{smirk_cat},
	\item
the particle was prepared to have the spin along the $x$ axis,
	\item
a measurement device is ready to measure the spin along the $z$ axis,
	\item
if the measurement finds the spin to be up, the cat gets milk \emoji{milk_glass}\emoji{smiley_cat}, and if it finds the spin to be down the cat gets broccoli \emoji{crying_cat_face}\emoji{broccoli}.
\end{itemize}

Then, the wavefunction evolves so that it spreads over two regions of the classical state space, corresponding to macrostates in which the cat got either milk or broccoli, as in Fig. \ref{fig:schrod:cat}.

\begin{ffig}
\begin{center}
\classicalManyToMany{-3}{3}{red}{100}{0}{1}{1}{1}
\captionof{figure}{{\schrod}'s cat as a classical cat, initially waiting to be fed \emoji{smirk_cat}, then getting either milk \emoji{milk_glass}\emoji{smiley_cat} or broccoli \emoji{crying_cat_face}\emoji{broccoli}.}
\label{fig:schrod:cat}
\end{center}
\end{ffig}

According to the classical theory of coexisting worlds from \ref{s:steps:wavefunction-complex}, the wavefunction consists of classical worlds that initially were in the initial macrostate containing a hungry cat, and then evolved either in the milk-macrostate, or in the broccoli-macrostate, as in Fig. \ref{fig:schrod:cat} (note that the worlds don't maintain their identities, but depicting this in the figure would've made it too complicated).
Since there are many classical worlds, they contain many instances of the cat. Any instance of the cat has to belong to only one of the classical worlds, so half of them receive milk, and the other half of them receive broccoli.
Because each world composing the wavefunction changes its microscopic identity over time, the cats change their microscopic identities as well.

\begin{center}
\textcolor{titcolor}{\noindent\rule{0.33\linewidth}{0.5pt}}
\end{center}

The same applies to complementary properties.
Bohr noticed that you can't measure the position and the momentum of the same particle at the same time. Each of these two measurements requires a different experimental setup, and no experimental setup can do both measurements simultaneously.
This is because a particle can't have at the same time both a definite position and a definite momentum. Some states of the particle have definite position and some states have definite momentum, but not both at once.
A quantum measurement finds the particle in one of the states in which the measured property has a definite value. That is, if we measure the position, the observed particle is found to have a definite position, and if we measure the momentum, it is found to have a definite momentum, as in Table \ref{table:bohr-complementarity}.

\begin{ftab}
\centering
\begin{tabular}{c|cc} 
					& \begin{tabular}{@{}c@{}}Definite\\position\end{tabular} &  \begin{tabular}{@{}c@{}}Definite\\momentum\end{tabular} \\
 \hline
 \begin{tabular}{@{}c@{}}Position\\measurement\end{tabular} & {\LARGE\textcolor{green!65!black}\cmark} & {\LARGE\textcolor{red}\xmark} \\
 \begin{tabular}{@{}c@{}}Momentum\\measurement\end{tabular} & {\LARGE\textcolor{red}\xmark} & {\LARGE\textcolor{green!65!black}\cmark} \\ 
\end{tabular}
\captionof{table}{Bohr complementarity: the observed system is found in a state in which the measured property has a definite value.}
\label{table:bohr-complementarity}
\end{ftab}

On the other hand, in classical mechanics, a point-particle always has both a definite position and a definite momentum.
So how can we then think that a classical reality, even upgraded as in \ref{s:steps:wavefunction-complex}, can explain this?
The reason is that in quantum mechanics this is no longer the case: only wavefunctions concentrated at a point have definite positions, and only wavefunctions spread across the entire space and waving like a sine function have definite momentum.
The same is true for the wavefunctional formulation, when the quantum fields are interpreted in terms of wavefunctions \citep{Hatfield2018QuantumFieldTheoryOfPointParticlesAndStrings}.
And, as explained in \ref{s:steps:wavefunction-complex}, the wavefunctional formulation can be understood as just a classical theory of coexisting worlds.

But even if it follows from this equivalence that quantum measurements can be accounted for by a classical theory as in \ref{s:steps:wavefunction-complex}, it would still be instructive to see how this is achieved.
The answer is simple:

\begin{frem}
There are possible classical worlds in which the experimental setup is so that the pointer indicates, upon measurement, a definite position, and the microscopic states of the observed system add-up to a wavefunction in which the observed system has a definite position. The same is true about momentum. The classical worlds containing a momentum measurement are distinct from those containing a position measurement.
\end{frem}

For the worlds in which the measured property of the observed particle doesn't have a definite value, the probability is zero.
So there are possible classical worlds for each combination, but the wavefunction contains only worlds in which the result of the measurement is consistent with the microstate, as in Table \ref{table:bohr-complementarity}.

Let's summarize and emphasize these essential observations:

\begin{frem}
The classical worlds making the wavefunction are such that the macrostate indicates what subsystem we observed, how it was prepared, what property was measured, and the result of the measurement. Microscopically, they are such that the wavefunction contains the observed system in a state in which the observed property has a definite value, the observed one.
But the microstates, the worlds, don't need to have a definite value for that property of the observed system, only the wavefunction obtained from them has to.
\end{frem}

\section{Built-in classicality for cats, tigers, and chairs}
\label{s:classicality}

But where in the wavefunction are the cats? More generally,

\begin{fquest}
If the world is quantum, why does it appear classical at the macroscopic level? Why does it present to us a manifest image containing objects like cats, tigers, and chairs?
\end{fquest}

A common proposal is that somehow classicality emerges from quantumness, that superpositions fade out as we move from the micro scale to the macro scale. Often a process called \emph{decoherence} is invoked \citep{JoosZeh1985EmergenceOfClassicalPropertiesThroughInteractionWithTheEnvironment,Zurek2003DecoherenceEinselectionAndTheQuantumOriginsOfTheClassical,Schlosshauer2007DecoherenceAndTheQuantumToClassicalTransition}.
Another explanation is that only the wavefunctions structured hierarchically into atoms, molecules, and so on, so that at higher levels of the hierarchy entanglement is absent, constitute worlds that appear to us macroscopically as being classical. That is, most entanglement is ``hidden'' within the atoms \citep{Vaidman2022WaveFunctionRealismAnd3Dimensions}.
Each of these explanations capture essential aspects of what happens.

In a reality of coexisting classical worlds as in \ref{s:steps:wavefunction-complex}, the answer is

\begin{frem}
We perceive the world as classical because we are in one of the classical worlds.
\end{frem}

As we've seen, even if reality is classical in this way, the evolution doesn't preserve the identity of the worlds, as in \ref{s:steps:identity}, and this is essential for our theory to recover quantum mechanics. The possibility to mix identities allows for quantum phenomena to exist.
The wavefunctional formulation of quantum field theory can be seen as a particular case of classical description in terms of coexisting classical worlds, so the dynamics of many classical worlds is just like quantum dynamics, and the quantum effects happen at the microscopic level. But at the macroscopic level we only experience one of the coexisting worlds, because each instance of us exists in only one of these classical worlds.

Now consider a situation in which the wavefunction, in the standard understanding, is structured so that entanglement is mostly hidden within the atoms and very little between atoms. At the macroscopic level, these atoms can be arranged like a cat \emoji{smirk_cat} and the other objects from Section \sref{s:indetermined}.
The world will then evolve into a superposition of worlds in which the cat gets milk \emoji{milk_glass}\emoji{smiley_cat} and respectively broccoli \emoji{crying_cat_face}\emoji{broccoli}.
If the spin was prepared not along the $x$ axis, but in a more general direction, the amplitudes of the two wavefunctions would be different.
Then we would have to postulate somehow that the probabilities are given by the two squared amplitudes.

In the classical theory of coexisting worlds, there are major differences.
Take for example the wavefunctional formulation seen as such a theory, where distinct classical configuration fields are represented by orthogonal vectors.
Then, even to describe a single atom you'd need a linear combination of basis vectors.
The same is true for the two worlds, in which the cat gets milk \emoji{milk_glass}\emoji{smiley_cat} and respectively broccoli \emoji{crying_cat_face}\emoji{broccoli} -- each of them requires an infinite number of basis vectors corresponding to classical field configurations \citep{Hatfield2018QuantumFieldTheoryOfPointParticlesAndStrings}.
But now we have a built-in probabilistic interpretation, because the two wavefunctionals are just classical probability distributions of classical field configurations. And we also have built-in classicality, because these worlds are classical.

However, if we could examine microscopically any of these classical worlds, we wouldn't find any atoms as described by {\schrod}.
But at the macro level they look just like the familiar objects that \emph{we know} are made of atoms.
This seems to be a contradiction, but it isn't, and it actually explains why things look quantum at micro level and classical at macro level.

\begin{frem}
Any experiment aiming to find the classical worlds at the microscopic level will only find superpositions of them, because these classical worlds don't maintain their identities.
\end{frem}

This is why, if we examine the cat's microscopic structure, we find atoms, and not classical field configurations.
And for this reason, even if reality consists of coexisting classical worlds, the usual formulation of quantum mechanics still provides a clearer description of what happens at the microscopic level.

\section{Can't we simply cut off the branches?}
\label{s:counterargument:cutt-branches}

If reality consists of coexisting worlds as in \ref{s:steps:wavefunction-complex}, then there are many instances of you in each of a huge (infinite!) number of worlds. Our baby steps led us to accept a new version of MWI, different from the usual ones as presented in \citep{Everett1957RelativeStateFormulationOfQuantumMechanics,Everett1973TheTheoryOfTheUniversalWaveFunction,Wallace2012TheEmergentMultiverseQuantumTheoryEverettInterpretation,SEP-Vaidman2021MWI}.
If the usual versions of MWI are ``extravagant'' by containing many worlds, the version presented here must be even more extravagant, as seen in Section \sref{s:indetermined}.

There is no known way to replace quantum mechanics by a single-world classical theory (as, for example, in \ref{s:steps:classical:deterministic:single:prob}), being it deterministic or not. Any attempt to do so has to overcome very strong restrictions following from Bell's theorem \citep{Bell64BellTheorem,Bell2004SpeakableUnspeakable} and the Kochen-Specker theorem \citep{KochenSpecker1967HiddenVariables,ConwayKochen2006TheFreeWillTheorem}.
There are still attempts to avoid these restrictions, but I will not discuss them here.

The next way to avoid the extravagance is the way usually applied to the standard versions of MWI, that is, whenever the world branches into many worlds, assume that only one of the branches, randomly chosen, is real, and cut-off all other branches. This is the standard way, and this cutting-off the branches is usually called \emph{wavefunction collapse} or \emph{state reduction}. It normally specifies the moment of collapse as being the moment of the measurement. A way to do this when the wavefunction spreads across more macrostates, as in Fig. \ref{fig:schrod:cat}, is presented in \citep{Stoica2020StandardQuantumMechanicsWithoutObservers}.
But whatever your preferred way to avoid many worlds, if it requires adding extra physical elements or modifying the {\schrod} equation, is it really less extravagant?

Einstein famously said ``God does not play dice''. Had he met Everett at Princeton and known about his theory, he could have seen that there is a way for this to be true. But then, noticing that Everett's theory requires many worlds, he could have objected that this is too extravagant,

\begin{fquote}
``God is not a big spender!''
\end{fquote}

Or if he wouldn't, at least many other people raised this objection, arguing that \emph{Ockham's razor}, the principle that we should always choose the simplest explanation, requires us to prefer a theory in which there is only one world.

However, I doubt that this imaginary Einstein would have this objection.
If he were a contemporary of Copernicus, would he object that a universe in which the stars are suns and may have their own planets, some of them with life, is too extravagant? I doubt this. This example of the Copernican theory as seeming to some of his contemporaries as extravagant was given by Tipler (\citealp{Tipler1986MWIofQM-InQuantumCosmology} p. 208):

\begin{fquote}
``one of the three most powerful arguments against the Copernican system was the existence of the vast useless interstellar void which the Copernican system required.''
\end{fquote}

By the same argument coming from misunderstanding Ockham's razor, one can claim that it's simpler that space is much smaller than it is, so that it contains our solar system, or maybe even just a small patch of space in which we exist, and the light from distant objects be the result of some vibration applied to the boundary of this patch.

We can also think that the world was created 6000 years ago, but in a state in which it seems 13.7 billion years older, to be consistent with our cosmological observations, the Carbon dating, and the fossils.
Or, why not, five minutes ago, but in the exact same state it was five minutes ago (\citealp{Russell1921AnalysisOfMind}, p. 159--160).

Or you could be a solipsist, because it would be much simpler if only you existed, and everyone else was a figment of your imagination. 

Why would God waste such an eternity and such an immense void, only for 15 minutes of life in a place lost in the immensity, when he could make the spacetime patch containing humanity with its entire history in a much smaller vat? And why even this, the Cartesian demon could ask while leading a rebellion against God on the basis that he could do it even cheaper, by running everything as a simulation, and even more so, just for you?

Reducing the entire universe to just a patch of spacetime containing the minimum necessary to explain this world doesn't make things cheaper. The laws of physics would have to be cut on the boundary, and replaced with new laws that mimic the appearance of the actual laws outside the boundary.

Similarly, is it really simpler to cut off the branches, instead of allowing them to be just what {\schrod}'s equation predicts? Isn't it like keeping only a small region of spacetime and arranging for the boundary conditions to ``impersonate'' the external region of the spacetime?

Anyway, suppose we go with this argument for simplicity. And suppose we liked the simplicity of the idea of making everything classical, as in this article. But then, since we need the wavefunction of the world anyway, if it consists of classical worlds, then we already accepted that there are many of them, coexisting until the next wavefunction collapse. And the wavefunction collapse would only cut off some of them, because the collapse in general results in another wavefunction, and not in a fully classical world.
So, if we already accepted that there are many classical worlds just to explain a single quantum world, 

\begin{frem}
why cut them off just to have a single macroscopic quantum world made of many microscopic classical worlds anyway?
\end{frem}

Nevertheless, it is not for me to judge what the reader considers to be a simpler explanation. We all came to quantum mechanics following a different path, and we framed the phenomena in different ways, which made us accept as plausible different views. So it makes sense that what to some may appear as a simpler way to understand quantum mechanics, we may seem as more complicated, and less plausible.
This is one of the reasons why I tried to build everything as much as possible from baby steps within a classical view of reality.

\section{``To be'' is sentiential}
\label{s:sentience}

There is an unexpected relation between consciousness and quantum mechanics. And I don't mean it in the way Heisenberg, von Neumann, and Wigner meant it, that the observers, when doing measurements, create elements of reality or collapse the wavefunction \citep{Heisenberg1958PhysicsAndPhilosophy,vonNeumann1955MathFoundationsQM,Wigner1967RemarksOnTheMindBodyProblemWignersFriend}, although there may be a connection with this.

Let's explore the idea that consciousness is just a computer program, a very sophisticated one, but still a computer program.
I don't really understand this claim, but many people seem to support it, so maybe they don't simply take the conventions behind the computations as objective, maybe they really mean something. Something that I missed \citep{Stoica2023DoesAComputerThinkIfNoOneIsAroundToSeeIt}?
Anyway, let's humor them and entertain this idea.
A computer is made of small parts that can be in two possible states. They can be for example capacitors, that can either be charged or discharged. There is in fact a continuum of states between ``charged'' and ``discharged'', and there are infinitely many possible states that count as ``charged'' or as ``discharged''.
So if a computer can be conscious, there are infinitely many worlds, differing very little, so that in each of these worlds the computer has the exact same conscious experience.
In other words, the computation that we associate to the physical process, the succession of physical states through which a computer goes, is in fact associated not to the microstates, but to a coarse graining of the state space, just like the coarse graining used to define macrostates in Figure \ref{fig:classical:deterministic:single:cert:macro}.

Now, consider instead of a classical world going through the succession of classical states underlying a computation, a wavefunction going through a succession of quantum states contained in the coarse graining associated to the computation. For example, if a memory holds the value of a bit, a superposition of classical states in which the memory holds the same value is consistent with the coarse graining. It follows then that there are many wavefunctions that appear as performing the same computation.

Now if the computer would have a sense of being, from which to derive a sense of self-location, it should be able to self-locate itself in superpositions of classical worlds as well.
And, if consciousness were a software running on the brain just like computer programs run on computers, this should be the case for our consciousness too.

But central to this article is the assumption that you can self-locate yourself only in one of the classical coexisting worlds, and not in any superposition of them, as explained in Section \sref{s:indetermined}.
If you could self-locate yourself in superpositions that only approximate the original classical state, then you'd have to count these possible states too, when deriving the probabilities, and this will mess-up the result (Proposition 1 in \citealp{Stoica2023TheRelationWavefunction3DSpaceMWILocalBeablesProbabilities}).
It follows that
\begin{frem}
You can have a sense of being only if you are in a classical state.
\end{frem}

The classical worlds form a basis in the vector space used in quantum mechanics to represent the quantum states of the world. These basis vectors are the only quantum states in which a sentient being can self-locate. General quantum states, being superpositions of classical states, support all instances of sentient beings existing in the classical states from the superposition.

\begin{ffig}
\begin{center}
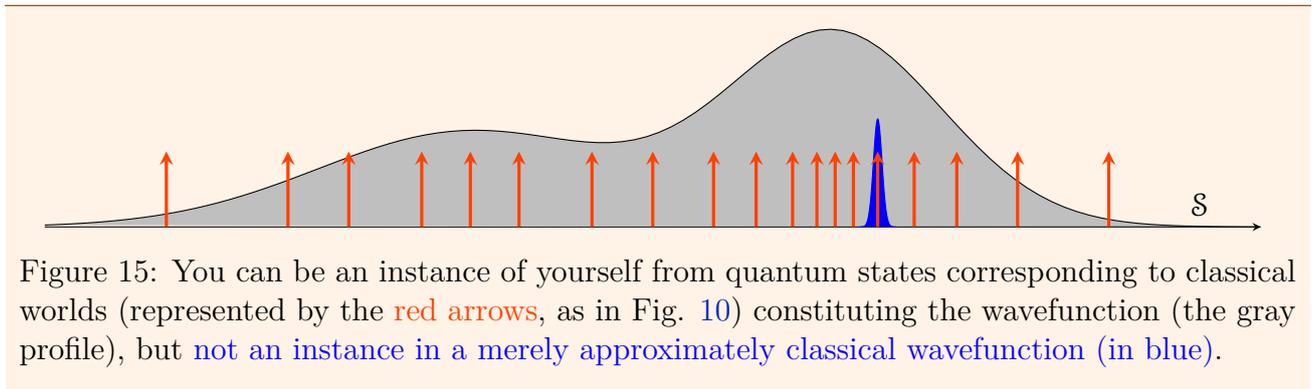

\coexistingdistribution{1}
\captionof{figure}{You can be an instance of yourself from quantum states corresponding to classical worlds (represented by the \textcolor{red!50!orange}{red arrows}, as in Fig. \ref{fig:classical:continuous-counting-a}) constituting the wavefunction (the gray profile), but \textcolor{blue}{not an instance in a merely approximately classical wavefunction (in blue)}.}
\label{fig:sentient-vectors}
\end{center}
\end{ffig}

Even if a quantum state approximates a human with a functioning brain, it doesn't endow it with a sense of self-location, a sense of being. It is like a \emph{philosophical zombie}, having all the right structures and processes, but insentient. 
\begin{frem}
Sentience is associated only with classical states. No approximation of such states is enough to yield sentience, because this would mess-up the self-location probabilities.
\end{frem}

Some theories of mind claim that, if we build a machine that imitates the brain's functionality close-enough, this would necessarily be conscious. The claim that consciousness reduces to computation is an example of such a theory of mind. Here we've seen that, no matter how closely a quantum state approximates a classical one, if it's not classical, it doesn't support sentience. Sentience is supported by the classical states that, superposed, give that quantum state. Therefore,  imitating closely the human brain is not sufficient to yield consciousness.
Or, at least, not in this version of MWI. For the reader interested in this subject, here are some proofs that are not constrained to this interpretation \citep{Stoica2023DoesAComputerThinkIfNoOneIsAroundToSeeIt,Stoica2023AskingPhysicsAboutPhysicalismZombiesAndConsciousness,Stoica2024SentientObserversAndTheOntologyOfSpacetime}.

\section{Conclusions}
\label{s:conclusions}

That's it, this is the theory of many coexisting classical worlds, which give us probabilities, classicality, complex numbers, and which works just well with quantum field theory. Whether or not the reader found the trip convincing and agrees with this proposal, there is more to be said. There are other foundational problems in quantum mechanics that I didn't discuss here, because I focused on the most direct implications. And there are technical details that I skipped, discussed in \citep{Stoica2022BornRuleQuantumProbabilityAsClassicalProbability,Stoica2022BackgroundFreedomLeadsToManyWorldsLocalBeablesProbabilities,Stoica2023TheRelationWavefunction3DSpaceMWILocalBeablesProbabilities}.

The point of this article was to explain a new interpretation of quantum mechanics, which is a version of MWI, and to collect feedback about it and about this way to explain it.

\addcontentsline{toc}{section}{\refname}


\end{document}